%% file: paper_ArXiv_v2.tex
\documentclass[12pt,a4paper, fullpage]{article}
\include{header}
\title{Singular Value-based Atmospheric Tomography with Fourier Domain Regularization (SAFR)}

\author{Lukas Weissinger\thanks{Johann Radon Institute Linz, Altenbergerstra{\ss}e 69, A-4040 Linz, Austria, (lukas.weissinger@ricam.oeaw.ac.at, bernadett.stadler@ricam.oeaw.ac.at, ronny.ramlau@ricam.oeaw.ac.at).}
\and Simon Hubmer\thanks{Johannes Kepler University Linz, Institute of Industrial Mathematics, Altenbergerstra{\ss}e 69, A-4040 Linz, Austria, (simon.hubmer@jku.at,ronny.ramlau@jku.at).}
\and Bernadett Stadler\footnotemark[1]
\and Ronny Ramlau\footnotemark[1]~\footnotemark[2]
}

\begin{document}
\maketitle
\begin{abstract}
Atmospheric tomography, the problem of reconstructing the atmospheric turbulence volume from wavefront sensor measurements, is an integral part of many adaptive optics systems. It is used to enhance the image quality of ground-based telescopes, such as for the Multiconjugate Adaptive Optics Relay For ELT Observations (MORFEO) instrument on the Extremely Large Telescope (ELT). To solve this problem, a singular value-type decomposition (SVTD) based approach has been proposed in previous research. In this paper, we focus on the numerical implementation of this SVTD approach, leading to the SVD-based Atmospheric Tomography with Fourier Domain Regularization Algorithm (SAFR), and investigate its performance for Multi-Conjugate Adaptive Optics (MCAO) systems. The key features of the SAFR algorithm are the utilization of the FFT and the pre-computation of computationally demanding parts. Together, this yields a fast algorithm with less memory requirements than commonly used Matrix Vector Multiplication (MVM) approaches. We evaluate the performance of SAFR regarding reconstruction quality and computational expense in numerical experiments using the simulation environment COMPASS, in which we use an MCAO setup resembling the physical parameters of the MORFEO instrument of the ELT.
\end{abstract}


%===========================================================================
%===========================================================================
% % % % % % % % % % % % % % % % % % % % % % %  
% % % % % % Section - Introduction  % % % % %
% % % % % % % % % % % % % % % % % % % % % % %  
\section{Introduction}\label{sec:introduction}

The image quality of ground-based telescopes suffers from temperature fluctuations in the atmosphere, which results in wavefront aberrations of the incoming light. These aberrations are typically corrected by an Adaptive Optics (AO) system to prevent an otherwise severe loss of image quality. Multi Conjugate Adaptive Optics (MCAO) systems, such as the MORFEO instrument of the ELT, aim for an aberration correction over a large field of view. For this, several wavefront sensors (WFSs) are used to estimate the resulting wavefront aberrations and correct for them using deformable mirrors (DMs) \cite{Roddier1999}. The WFSs use the incoming light from either natural guide stars (NGS) or artificial laser guide stars (LGS), which are created by powerful lasers in the sodium layer of the atmosphere, that act as reference objects.

The calculation of optimal mirror shapes for the DMs is commonly done by inverting the so called interaction matrix. This matrix relates DM commands to WFS measurements. The multiplication of the WFS measurements with this inverse matrix is often referred to as the Matrix Vector Multiplication (MVM) method.
Since the atmosphere changes rapidly, the mirror shapes have to be updated at a rate of about 500-1000 Hz \cite{Capasso_2024}. To meet this strict time requirements, MVM methods can leverage parallel computations. However, the computational demand of a matrix vector multiplication of $\mathcal{O}(N^2)$, where $N$ denotes the number of sensor measurements, is still high. Therefore, as telescope sizes increase, the demand for efficient algorithms has grown as well. Hence, as an alternative to MVM methods, three-step approaches \cite{Saxenhuber_2016} have been proposed, where the full AO-task is split into the $3$ sub-problems: 
\begin{enumerate}
    \item \textbf{Wavefront reconstruction} from wavefront sensor measurements. % $s$.
    \item \textbf{Reconstruction of atmospheric turbulence} from wavefronts.
     \item \textbf{Computation of deformable mirror shapes} from turbulence layers.
\end{enumerate}

For solving the first of these sub-problems, different wavefront reconstruction methods have been proposed \cite{Gilles2002, Poyneer_2002, Quiros_Pacheco_2010, Shatokhina_2013, Thiebaut_Tallon_2010, Vogel_Yang_2006a}. For the numerical experiments considered in Section~\ref{sec:numerics}, we use the Cumulative Reconstructor with Domain Decomposition (CuReD) \cite{Rosensteiner2011a, Rosensteiner2012}, a reliable and efficient method scaling at $\mathcal{O}(N)$.  

\begin{figure}[ht!]
	\centering
	%\includegraphics[width=0.45\textwidth]{Figures/AO_Tomo_full}
	%\qquad
	%\includegraphics[width=0.45\textwidth]{Figures/AO_Tomo_cone}
    \includegraphics[width=\textwidth]{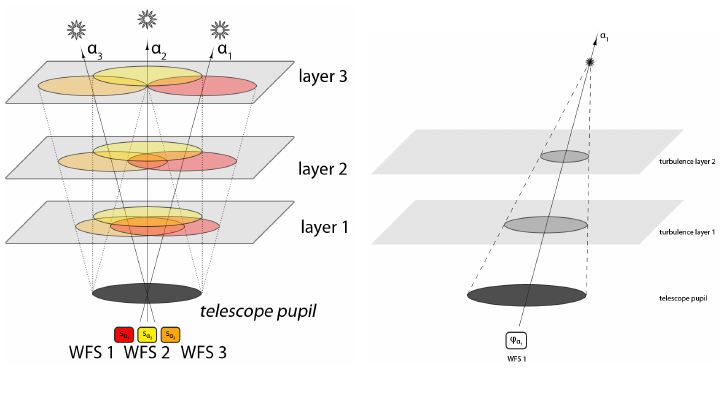}
	\caption{Illustration of the atmospheric tomography problem with three turbulence layers, NGSs and corresponding WFSs (left). Light stemming from a single LGS is influenced by the cone effect (right). Images taken with permission from~\cite{Yudytskiy_2014}.}
	\label{fig_AO_Tomo}
\end{figure}

In this paper, we focus on the second sub-problem of the atmospheric tomography problem, i.e., the reconstruction of the turbulence volume above the telescope from the reconstructed wavefronts. Atmospheric tomography is a \textit{limited-angle} tomography problem, since only a small number of guide stars with a small angle of separation are available. As such, it is severely ill-posed and thus practically infeasible to solve without additional restrictions \cite{Davison_1983, Natterer_2001,Ramlau_Stadler_2025}. Hence, one commonly assumes that the atmosphere consists of a finite number of (infinitely) thin turbulent layers located at predefined heights, and aims to reconstruct this layered atmospheric turbulence profile. An example of this problem for the case of three layers and three (natural) guide stars with corresponding WFSs is shown in Fig.~\ref{fig_AO_Tomo} (left).

The third sub-problem is usually referred to as mirror fitting and can be solved using, e.g., Kaczmarz iteration \cite{Rosensteiner_Ramlau_2013} or the conjugate gradient (CG) method \cite{Fusco_Conan_Rousset_Mugnier_Michau_2001,Yudytskiy_2014}. However, this step can be omitted if the layer heights are chosen such that the light emitted from a certain turbulent layer is exactly focused in the corresponding DM. This special choice of the mirror position along the optical axis is typically referred to as conjugation 
and allows the DM shapes to be obtained directly from the reconstructed layers.

The atmospheric tomography problem has attracted considerable attention in the past. Among the many proposed reconstruction methods we mention the minimum mean square error method \cite{Fusco_Conan_Rousset_Mugnier_Michau_2001}, CG type iterative reconstruction methods with suitable preconditioning \cite{Ellerbroek_Gilles_Vogel_2003,Gilles_Ellerbroeck_2008,Gilles_Ellerbroek_Vogel_2003,Vogel_Yang_2006,Yang_Vogel_Ellerbroek_2006}, Fourier transform-based methods \cite{Tokovinin,Gavel_2004}, the Fractal Iterative Method (FrIM) \cite{Tallon_TallonBosc_Bechet_Momey_Fradin_Thiebaut_2010,Tallon_Bechet_TallonBosc_Louarn_Thiebaut_Clare_Marchetti_2012,Thiebaut_Tallon_2010}, the Finite Element Wavelet Hybrid Algorithm (FEWHA) \cite{Stadler2021,Stadler2020,Yudytskiy_2014,Yudytskiy_Helin_Ramlau_2013,Yudytskiy_Helin_Ramlau_2014}, as well as a Kaczmarz iteration \cite{Ramlau_Rosensteiner_2012, Rosensteiner_Ramlau_2013}. Recently, a deep-learning-based approach has been considered in~\cite{Zhang_2024}. For further methods as well as important practical considerations see also~\cite{Ellerbroek_Gilles_Vogel_2002,Gilles_Ellerbroek_Vogel_2007,Gilles_Ellerbroek_Vogel_2002,Poettinger_Ramlau_Auzinger_2019,Raffetseder_Ramlau_Yudytskiy_2016,Ramlau_Obereder_Rosensteiner_Saxenhuber_2014,Saxenhuber_Ramlau_2016} and the references therein. FEWHA is an iterative reconstruction algorithm which uses a dual-domain discretization strategy into wavelet and bilinear basis functions leading to sparse operators. A matrix-free representation of all involved operators makes FEWHA very fast and enables on-the-fly system updates whenever parameters at the telescope or in the atmosphere change \cite{Stadler2021,Stadler2020}. The resulting sparse system is solved using the CG method with a Jacobi preconditioner \cite{Yudytskiy_Helin_Ramlau_2013} and an augmented Krylov subspace method \cite{Stadler2021,Stadler2020}. We use FEWHA as a benchmark algorithm, since it has been shown that it leads to an excellent reconstruction quality compared to standard MVM approaches for the MORFEO instrument, while also being computationally efficient \cite{Stadler2021,Stadler2022}.

To provide theoretical insights into the ill-posedness of the atmospheric tomography problem, a singular value decomposition (SVD) has been proposed in~\cite{Neubauer_Ramlau_2017}. This SVD has only been derived for square telescope apertures in an NGS-only setting, which limits its practical applicability. In~\cite{Hubmer_Ramlau_2020}, the SVD has been extended to an LGS-only setting. Moreover, a frame-decomposition was proposed, which provides an SVD-like decomposition of the atmospheric tomography operator allowing for general aperture shapes and a mixture of NGS and LGS. The main drawback of the frame-decomposition is that it only provides an approximate solution to the atmospheric tomography problem; cf.~\cite{ Hubmer_Ramlau_2021_01,Weissinger2021,WeissingerHubmerStadlerRamlau_2025}. Finally, in~\cite{WeissingerHubmerStadlerRamlau_2025}, the SVD of the atmospheric tomography operator was further extended to incorporated atmosphere statistics, leading to a so-called singular value-type decomposition (SVTD), which forms the basis of the SAFR algorithm considered here.

Motivated by the excellent performance of the SVTD method in~\cite{WeissingerHubmerStadlerRamlau_2025}, in this paper we particularly study SAFR in the context of a three-step method for MCAO systems. We give a detailed description of the algorithm and conduct a performance analysis for the MORFEO instrument using the simulation environment COMPASS \cite{Ferreira_2018}. Moreover, we provide a detailed study
of the quality as well as the computational complexity of SAFR, and propose a strategy to overcome the limitation that the classic SVD-approach can only be used in an LGS or NGS-only setting.

The paper is organized as follows: The atmospheric tomography problem and the related operator are defined in Sec.~\ref{sec:tomo}. Then, we recall the SVD as a method to solve the atmospheric tomography problem and derive the SAFR algorithm in Sec.~\ref{sec:SAFR}. The quality and computational performance of the algorithm within an AO system compared against FEWHA and an MVM-like method is demonstrated by numerical simulations in
Sec.~\ref{sec:numerics}. In Sec.~\ref{sec:conclusion}, we present our final conclusions.

% % % % % % % % % % % % % % % % % % % % % % % % %
% Section - The Atmospheric Tomography Operator %
% % % % % % % % % % % % % % % % % % % % % % % % %
\section{Atmospheric Tomography in Three-step Adaptive Optics}\label{sec:tomo}

In this section, we introduce the mathematical formulation of the atmospheric tomography problem using a layered structure of the turbulence \cite{Ellerbroek_Vogel_2009,Fusco_Conan_Rousset_Mugnier_Michau_2001}.
Let $h_\LGS$ denote the height of the sodium layer in the atmosphere in which LGSs are created (approximately $90$ km). We assume that there are $L$ atmospheric layers, i.e., planes parallel to the telescope aperture $\Omega_A$, located at distinct heights $h_{\ell} \in [0,h_\LGS)$ for $\ell = 1 \,, \dots \,, L$. The heights are assumed to be in ascending order. We consider $G$ different guide stars with corresponding direction vectors $\alpha_g = (\alpha_g^x,\alpha_g^y) \in \R^2$ for $g = 1\ ,,\dots \,, G$. The vectors $\alpha_g$ are chosen such that, seen from the center of the telescope aperture, the vectors $(\alpha_g^x,\alpha_g^y,1) \in \R^3$ point at the corresponding guide stars.
Here, we consider the NGS-only or LGS-only case and correspondingly define
    \begin{equation*}
        c_{\ell}:=
        \begin{cases}
        1\,,\quad & \text{NGS-only setting}\,,
        \\ 1- h_\ell/h_{\LGS} \,, \quad & \text{LGS-only setting} \,.
        \end{cases}
    \end{equation*}
 The coefficients $c_{\ell}$ model the cone effect for LGS as illustrated in Fig.~\ref{fig_AO_Tomo} (right). Note that since $h_\ell < h_{LGS}$ for all $\ell=1,\ldots,L$, we have $c_{\ell} \in (0,1]$.

\begin{figure}[ht!]
    \centering
    \includegraphics[width=0.6\linewidth]{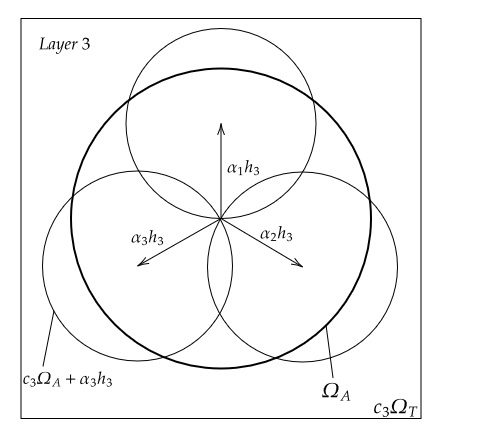}
    \caption{Illustration of the domains $c_\ell\Omega_T$ and $c_\ell\Omega_A+\alpha_g h_\ell$ for $\ell = 3$, the highest layer in the atmospheric tomography setting with $3$ layers and $3$ guide stars schematically depicted in Figure~\ref{fig_AO_Tomo}.}
    \label{fig:domains2D}
\end{figure}

Furthermore, we define the square domain $\Omega_T=[-T,T]^2 \subset \R^2$, with $T$ chosen sufficiently large, such that 
\begin{equation*}
    c_\ell\Omega_A+\alpha_g h_\ell\subset c_\ell \Omega_T, \qquad \forall g={1,\ldots,G},\quad \forall \ell={1,\ldots,L}\,,
\end{equation*}
where $c_\ell\Omega_T := [c_\ell T,c_\ell T]^2.$
I.e., the square domain $c_\ell\Omega_T$ fits both the telescope aperture as well as its copies - shifted and scaled depending on the guide star location - in each layer; see Figure~\ref{fig:domains2D}.

For the incoming wavefronts $\varphi=(\varphi_g)_{g=1,\ldots,G}$ we assume that $\varphi_g\in L_2(\Omega_T),$ i.e., that they are square integrable, and that they are (component-wise) periodic.
To incorporate atmosphere statistics into the model, we make use of periodic Sobolev spaces $H^\beta\left(c_{\ell} \Omega_T,\gamma_\ell\right)$, where $\beta\in\R^+$ denotes a smoothness parameter. The weights $\gamma_\ell$ are used to place a higher emphasis on layers on which a strong turbulence is expected, and are normalized such that they satisfy $\sum_{\ell=1}^L \gamma_\ell=1$. In practice, they are known quantities derived from previously measured turbulence profiles \cite{RODDIER1981,Saxenhuber17}. Note that this Sobolev-space setting is motivated by a result of Kolmogorov \cite{Kolmogorov}, due to which a typical atmospheric turbulence layer is expected to satisfy $\beta = 11/6$. This assumption is commonly used for both the simulation of atmospheric turbulence profiles \cite{Octopus,octopus06} as well as in modern reconstruction algorithms for atmospheric tomography; see e.g.~\cite{Auzinger_2017,Eslitzbichler2013,Yudytskiy_Helin_Ramlau_2014}.

Let $\phi=(\phi_\ell)_{\ell=1,\ldots,L}$ with $\phi_\ell \in H^\beta\left(c_{\ell} \Omega_T,\gamma_\ell\right)$, for given turbulence weights $(\gamma_\ell)_{\ell=1,\dots,L}$ and $\beta \in \R^+$. These $\phi_\ell$ denote the turbulence layers, i.e., functions describing the refractive index variations related to temperature fluctuations within the atmosphere at different layers. Then the atmospheric tomography operator $A=(A_g)_{g=1,\ldots,G}$ is defined as \cite{Fusco_Conan_Rousset_Mugnier_Michau_2001}
    \begin{equation}\label{defAperonly}
        \qquad(A_g \phi) (r):= \sum_{\ell=1}^L\phi_{\ell}(c_{\ell} r + \, \alpha_g h_{\ell})  \,.
    \end{equation}
The operator $A$, as defined in \eqref{defAperonly}, sums up the contributions of each turbulence layer in the direction of the guide stars. 
Note that so far we considered square domains, while a telescope aperture is typically circular. However, functions known only on the telescope apertures can be extended by zero to the full square.

The effect of this extension was studied in~\cite{Gerth_Hahn_Ramlau_2015}, where it was found to produce only very minor errors in the reconstructions along the boundaries of the domain $\Omega_A$.
Note that this extension also ensures periodicity, which is therefore no practical restriction of our model.

% % % % % % % % % % % % % % % % % % % % % % %  
% % % % % % Section - SVD % % % % % % % % % %
% % % % % % % % % % % % % % % % % % % % % % %  
\section{The Singular Value-based Atmospheric Tomography
with Fourier Domain Regularization Algorithm}\label{sec:SAFR}
In this section, we recall the SVD for the atmospheric tomography operator $A$ as first proposed in~\cite{Neubauer_Ramlau_2017} and extended in~\cite{Hubmer_Ramlau_2020,WeissingerHubmerStadlerRamlau_2025} (where it is then called SVTD). Furthermore, we introduce the SAFR algorithm and focus on the computational aspects of this method. 

\subsection{Mathematical description}
It is known \cite{Hubmer_Ramlau_2020,Neubauer_Ramlau_2017,WeissingerHubmerStadlerRamlau_2025} that the least-squares solution $\phi^\dagger$ of the equation $A\phi=\varphi$ with $A$ as defined in \eqref{defAperonly} (i.e., LGS-only or NGS-only case) is given by 
    \begin{equation*}\label{Aperdagger}(\phi^\dagger)_\ell(x,y):=\sum_{j,k\in\mathbb{Z}}\left(A^\dagger_{jk}\varphi_{jk}\right)_\ell w^{(\beta)}_{jk,\ell}(x,y) \,,
\end{equation*}
where
\begin{equation*}
    \varphi_{jk} = \kl{\spr{\varphi_g,w_{jk}}_{L_2(\Omega_T)}}_{g=1}^G \in\mathbb{C}^G \,,
\end{equation*} 
and $A_{jk}^\dagger$ denotes the pseudo-inverse of the matrix $A_{jk}\in\mathbb{C}^{G\times L}$, which is defined by
\begin{equation*}\label{Ajkdef}
    A_{jk} := \kl{ (2T)w^{(\beta)}_{jk,\ell}(\alpha_g^x h_\ell,\alpha_g^y h_\ell) }_{g,\ell=1}^{G,L} \,.
\end{equation*}
Here, we use the functions 
\begin{equation*}\label{def_wjk_wjkl}
    w_{jk,\ell}^{(\beta)}(x,y) :=\left(1+\tau_{\ell,T}|(j,k)|^2\right)^{-\beta/2}\frac{\gamma_\ell^{1/2}}{c_{\ell}}w_{jk}((x,y)/c_{\ell})\,, \quad \tau_{\ell,T}=\pi^2(c_\ell T)^{-2} \,,
\end{equation*}
and
\begin{equation*}
    w_{jk}(x,y):=\frac{1}{2T}e^{i\omega(jx+ky)} \,, \quad \omega=\frac{\pi}{T} \,,
\end{equation*}
which form orthonormal bases over the spaces $H^\beta(c_{\ell} \Omega_T,\gamma_{\ell})$ and $L_2(\Omega_T)$, respectively.

Mathematically, the atmospheric tomography problem is ill-posed \cite{Engl_Hanke_Neubauer_1996, Ramlau_Stadler_2025}, and thus some form of regularization is required in case of noisy data $\varphi^\delta$.
For this, consider the SVD of the matrix $A_{jk}$, i.e., $A_{jk}=U\Sigma V^H.$ The inversion of $A_{jk}$ is ill-conditioned \cite{Neubauer_Ramlau_2017}, and thus, we obtain a regularized inverse of $A_{jk}$ via $\mathcal{R}_{\alpha}(A_{jk})=Vg_\alpha(\Sigma^T) U^H, $ where $g_\alpha:\R\to\R$  denotes a regularizing filter (cf.~\cite{Engl_Hanke_Neubauer_1996}) which acts component-wise on the non-zero elements of $\Sigma^T.$ A regularized solution for the atmospheric tomography problem can then be obtained by
\begin{equation}\label{SVD_reg}
    \left(\phi_\alpha^\delta\right)_\ell (x,y):=\sum_{j,k\in\mathbb{Z}}\left(\mathcal{R}_{\alpha}(A_{jk})\varphi^\delta_{jk}\right)_\ell w^{(\beta)}_{jk,\ell}(x,y) \,.
\end{equation}
In order for this approach to approximate the least-squares solution $\phi^\dagger$, the filter $g_\alpha$ needs to be chosen appropriately, approaching $s\mapsto 1/s$ as $\alpha \to 0$. In our numerical experiments, we use a Tikhonov type filter of the form $g^{\text{Tikh}}_\alpha(s) :=s/(s^2+\alpha)$, which generally led to the best performance in our numerical experiments, together with a suitably selected regularization parameter $\alpha$. Note that throughout our analysis, we used one fixed regularization parameter for the inversion of all matrices $A_{jk}$. This approach could be further refined by using independent regularization parameters for every matrix $A_{jk}.$ Furthermore, other filter functions or truncation can be used as well \cite{Engl_Hanke_Neubauer_1996}. Note that the practical necessity of truncating the infinite sums in \eqref{SVD_reg} at a finite index provides an additional (but by itself in general insufficient) regularization effect in any implementation. Furthermore, note that FEWHA is also based on a specific Tikhonov regularization approach \cite{Stadler2021}.

Finally, note that together with the coefficients 
\begin{equation}\label{def_bjkl}
    b_{jk,\ell}:=\left(1+\tau_{\ell,T}|(j,k)|^2\right)^{-\beta/2}\left(\mathcal{R}_{\alpha}(A_{jk})\varphi^\delta_{jk}\right)_\ell \,,
\end{equation} 
and using the definition of $w^{(\beta)}_{jk,\ell}$, we can equivalently rewrite \eqref{SVD_reg} as
\begin{equation}\label{SVD_reg_algo}
    \begin{aligned}
        \left(\phi_\alpha^\delta\right)_\ell  (x,y)
        &=\sum_{j,k\in\mathbb{Z}}\left(\mathcal{R}_{\alpha}(A_{jk})\varphi^\delta_{jk}\right)_\ell w^{(\beta)}_{jk,\ell}(x,y) 
        \\
        &=
        \sum_{j,k\in\mathbb{Z}}\left(1+\tau_{\ell,T}|(j,k)|^2\right)^{-\beta/2}\left(\mathcal{R}_{\alpha}(A_{jk})\varphi^\delta_{jk}\right)_\ell w^{(0)}_{jk,\ell}(x,y) \\
        &=\sum_{j,k\in\mathbb{Z}}b_{jk,\ell} w_{jk,\ell}^{(0)}(x,y)\,,
    \end{aligned}
\end{equation}
which forms the basis of the numerical implementation considered below.

\subsection{Numerical Implementation}

\begin{figure}
    \centering
    \includegraphics[width=0.6\linewidth]{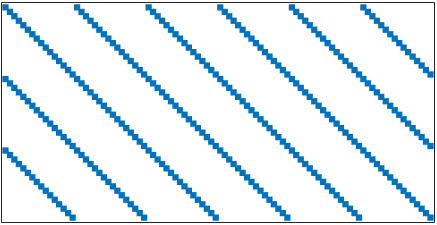}
    \caption{Structure of the matrix $B$, for $L=3$, $G=6$. Blue dots indicate nonzero entries of $B$.}
    \label{fig:visB}
\end{figure}

In this section, we consider the numerical implementation of the reconstruction formula \eqref{SVD_reg_algo}, which leads to the definition of the SAFR algorithm. Details for the mathematical derivation can be found in Appendix~\ref{appendix}. First, we consider equidistant grids for discretization on the square domains $c_\ell \Omega_T$ and $\Omega_T$, respectively, using a separate grid for each layer.
 I.e, we use the discretization
    \begin{equation*}
        (x_p^\ell,y_q^\ell)=(-c_{\ell}T+2c_{\ell}T \cdot p/M,-c_{\ell}T+2c_{\ell}T \cdot q/M),\qquad  \text{for} \quad p,q=0,\ldots,M-1 \,,
    \end{equation*}
and
    \begin{equation*}
        (x_p,y_q)=(-T+2T \cdot p/M,-T+2T \cdot q/M),\qquad  \text{for} \quad p,q=0,\ldots,M-1 \,.
    \end{equation*}
The discretized version of \eqref{SVD_reg_algo} is now the computation of $\Phi^{\alpha,\delta}=(\Phi^{\alpha,\delta}_\ell)_{\ell=1,\ldots,L}$, where
    \begin{equation}\label{def_discrete_turbulence}
        \Phi^{\alpha,\delta}_\ell\approx \left((\phi^\delta_\alpha)_\ell(x_p^\ell,y_q^\ell)\right)_{p,q=0}^{M-1}\in\R^{M\times M}
    \end{equation}
from the given (discretized) wavefront data $\Psi^\delta=(\Psi^\delta_g)_{g=1,\ldots,G}$, with
    \begin{equation}\label{def_discrete_wavefront}
        \Psi^\delta_g:=(\varphi^\delta_g(x_p,y_q))_{p,q=0}^{M-1}\in\R^{M\times M}.
    \end{equation}
%%%%%%%%derivation
Using the girds introduced above, the coefficients $\varphi_{jk}^\delta$ and the outer sums can be approximated by the two-dimensional Discrete Fourier Transform (DFT), which is defined by
    \begin{equation*}
        \texttt{DFT2}(X)_{j+1,k+1}= \sum_{p,q=0}^{M-1} e^{-2\pi i (jp+kq)/M} X_{p+1,q+1}\,, \qquad  \text{for} \quad j,k=0,\ldots,M-1 \,,
    \end{equation*}
and its inverse
    \begin{equation*}
        \texttt{IDFT2}(Y)_{p+1,q+1}=\frac{1}{M^2} \sum_{j,k=0}^{M-1}e^{2\pi i (jp+kq)/M} Y_{j+1,k+1} \,, \qquad  \text{for} \quad p,q=0,\ldots,M-1 \,,
    \end{equation*}
for $X,Y \in \C^{M\times M}$. Note that the DFT2 can be efficiently implemented via the two-dimensional Fast Fourier Transform (\texttt{fft2}). 
To align a symmetric truncation of the infinite sum in \eqref{SVD_reg_algo} with the indices used in the definition of the DFT, we define the shifted indices
    \begin{equation*}
        n(j):= \begin{cases}
        j\,, \qquad &j \leq m\,, \\
        j-M\,, \quad & m < j < M\,.
        \end{cases}
    \end{equation*}
With these preliminaries, we can then show (see Appendix~\ref{appendix} for details) that
    \begin{equation}\label{idft_approximation}
        \left(\phi_\alpha^\delta\right)_{\ell}(x_p^\ell,y_q^\ell)\approx
        \frac{\gamma_\ell^{1/2}M^2}{2c_\ell T}\texttt{IDFT2}(b_{\ell})_{p+1,q+1} \;,
    \end{equation}
with the vector $b_{\ell} := (b_{n(j)n(k),\ell})_{j,k=0}^{M-1}$, where the coefficients $b_{n(j)n(k),\ell}$ satisfy
    \begin{equation}\label{bjkl_approximation}
        b_{n(j)n(k),\ell} \approx \frac{2c_\ell T}{\gamma_\ell^{1/2}M^2}\sum_{g=1}^G b_{n(j)n(k),\ell g} \cdot \texttt{DFT2}(\Psi_g^\delta)_{j+1,k+1}\;.
    \end{equation}
with
    \begin{equation*}
        b_{jk,\ell g}:=\frac{\gamma_\ell^{1/2}}{c_\ell}\left(1+\tau_{\ell,T}|(jk)|^{2}\right)^{-\beta/2}\left(\mathcal{R}_{\alpha}(A_{jk})\right)_{\ell g}\,.
    \end{equation*}
The terms ${\gamma_\ell^{1/2}M^2}/(2c_\ell T)$ and ${2c_\ell T}/(\gamma_\ell^{1/2}M^2)$ in \eqref{idft_approximation} and \eqref{bjkl_approximation} cancel out in the final algorithm.
Furthermore, we define a sparse matrix $B\in \C^{LM^2\times GM^2}$ with entries
    \begin{equation}\label{sparseM}
        (B)_{u(j,k,\ell),v(\Tilde{j},\Tilde{k},g)}=\begin{cases}
        b_{n(j)n(k),\ell g}\,,\quad &j=\Tilde{j} \text{ and }k=\Tilde{k},\\
        0,\quad &\text{else.} 
        \end{cases}
    \end{equation}
Here, the one-to-one mappings 
    \begin{equation*}
    \begin{aligned}
        &u(j,k,\ell)=(\ell-1) \cdot M^2+j \cdot M+k+1\,,\\
        &v(\Tilde{j},\Tilde{k},g)=(g-1) \cdot M^2+\Tilde{j} \cdot M+\Tilde{k}+1\,,
    \end{aligned}
    \end{equation*} 
are used to convert the three-dimensional indices into linear indices. In particular, $B$ is non-zero only in $LGM^2$ of its total $LGM^4$ entries, see Fig.~\ref{fig:visB}.
Combining \eqref{idft_approximation} and \eqref{bjkl_approximation}, we can now define the SAFR algorithm via its pseudo-code given in Algorithm~\ref{alg:SAFR_hrt} (hard real-time computations) and Algorithm~\ref{alg:SAFR_srt} (soft real-time computations). Note that the hard real-time computations of SAFR are related to the computation of the mirror commands (via the turbulence layers) from sensor measurements, which have to be done throughout the AO observation, while the soft real-time computations are related to the pre-computation of the involved matrices, which only have to be (re)done whenever certain parameters at the telescope or in the atmosphere change.

\begin{algorithm}[ht!]
    \caption{SAFR - Hard real-time computations}\label{alg:SAFR_hrt}
    \begin{algorithmic}[1]
        \setstretch{1.4}
        \Require $\Psi^\delta_1,\ldots,\Psi^\delta_G\in \R^{M\times M},\, B$ from Algorithm~\ref{alg:SAFR_srt}; cf.~\eqref{sparseM}
        \For{$g=1,\ldots,G$}
        \State $d_g=\texttt{fft2}(\Psi^\delta_g)$ 
        \EndFor
        \For{$j,k=0,\ldots,M-1,\, \ell=1,\ldots,L$} 
        \State $\left(b_{\ell}\right)_{j+1,k+1}=\displaystyle\sum_{g=1}^G b_{n(j)n(k),\ell g}\cdot(d_{g})_{j+1,k+1}$
        \EndFor
        \For{$\ell=1,\ldots,L$}
        \State $(\Phi_\alpha^\delta)_\ell=\texttt{ifft2}(b_\ell)$
        \EndFor 
        \State\Return $(\Phi_\alpha^\delta)_1,\ldots,(\Phi_\alpha^\delta)_L \in\R^{M\times M}$
    \end{algorithmic}
\end{algorithm}

\begin{algorithm}[ht!]
    \caption{SAFR - Soft real-time computations}\label{alg:SAFR_srt}
    \begin{algorithmic}[1]
        \setstretch{1.4}
        \Require $\alpha\in\R^+,\, \beta\in\R^+$
        \For{$j,k=0,\ldots,M-1$} 
        \State $[U_{n(j)n(k)},\Sigma_{n(j)n(k)},V_{n(j)n(k)}]=\texttt{SVD}(A_{n(j)n(k)})$
        \For{$g=1,\ldots,G, \ell=1,\ldots,L$}
        \State 
            $b_{n(j)n(k),\ell g}=\frac{\gamma_\ell^{1/2}}{c_\ell}\left(1+\tau_{\ell,T}|(n(j),n(k))|^{2}\right)^{-\beta/2}$
        \Statex
            $\hfill\times \displaystyle\sum_{r=1}^{\text{rank}(A_{n(j)n(k)})}\frac{\left(\Sigma_{n(j)n(k)}\right)_{r,r}}{\left(\Sigma_{n(j)n(k)}\right)^2_{r,r}+\alpha}\left({U}^{H}_{n(j)n(k)}\right)_{r,g}\left(V_{n(j)n(k)}\right)_{r,\ell}$    
        \EndFor
        \EndFor
        \State\Return $B$ as in \eqref{sparseM}
    \end{algorithmic}
\end{algorithm}

Note that the SAFR algorithm can also be interpreted as a certain type of MVM method. To see this, we start by defining the block-diagonal matrices
\begin{equation*}
    \texttt{IDFT2}_L:=\underbrace{\begin{pmatrix}
        \texttt{IDFT2} & & \\
        & \ddots & \\
        & & \texttt{IDFT2}
    \end{pmatrix}}_{\text{L times}}\,,\qquad
    \texttt{DFT2}_G:=\underbrace{\begin{pmatrix}
        \texttt{DFT2} & & \\
        & \ddots & \\
        & & \texttt{DFT2}
    \end{pmatrix}}_{\text{G times}} \,,
\end{equation*}
where the blocks $\texttt{(I)DFT2}$ denote the two-dimensional (inverse) discrete Fourier transform in matrix form. Together with the matrix $B$ given in \eqref{sparseM}, we then obtain a matrix representation of the SAFR Algorithm~\ref{alg:SAFR_hrt}: The reconstructed atmospheric turbulence layers $\Phi^{\alpha,\delta}=(\Phi^{\alpha,\delta}_\ell)_{\ell=1,\ldots,L}$ as given in \eqref{def_discrete_turbulence} can be written as
\begin{equation*}\label{eq:SAFR_matrixrep}
    \Phi^{\alpha,\delta} = \texttt{IDFT2}_L \cdot B \cdot \texttt{DFT2}_G \cdot \Psi^\delta\,,
\end{equation*}
for known wavefront data $\Psi^\delta =(\Psi^\delta_g)_{g=1,\ldots,G}$ as defined in \eqref{def_discrete_wavefront}.

\subsection{The SAFR Algorithm in Three-step AO}

The SAFR algorithm is only concerned with the atmospheric tomography step, and is thus independent of the algorithms used for wavefront reconstruction, mirror fitting, and the control strategy.
Now, we discuss how we incorporate SAFR in a full AO system, where deformable mirror shapes have to be calculated in every time step. For this, we consider the availability of closed loop Shack-Hartmann (SH) WFS measurement data. Closed loop refers to residual wavefronts, i.e., already corrected with the mirror shapes from the previous time step. Furthermore, in realistic MCAO settings, LGS and NGS are mixed, due to the tip-tilt uncertainty of LGS \cite{Gilles_Ellerbroeck_2008}. However, the SAFR Algorithm itself is only viable for setups where LGS and NGS are not mixed. To still use a three-step method with SAFR in such cases, we use a split tomography approach where NGS are used only for the reconstruction of the tip-tilt component of the atmosphere (referred to as tip-tilt stars (TTS)), while the tip-tilt component from the LGS is completely removed \cite{Gilles_Ellerbroeck_2008}. Then, the TTS and LGS systems can be treated separately by SAFR and, due to the reconstruction of the distinct components, the reconstructed atmospheric turbulence layers can be added. In that case, we denote $G=G_\LGS+G_{\text{TTS}}$, such that $(\Psi_1,\ldots,\Psi_{G_\LGS})$ correspond to wavefronts stemming from LGS and $(\Psi_{\LGS+1},\ldots,\Psi_{G})$ correspond to wavefronts stemming from TTS.

In our numerical experiments, we use a specific three-step AO system, which is shown as pseudo-code in Algorithm~\ref{alg:3step}. We use the CuReD algorithm for wavefront reconstruction and choose the atmospheric turbulence layers such that the mirror fitting step can be omitted. To increase stability and robustness, we use a  pseudo open loop control (POLC) approach \cite{Ellerbroek2003b}, in which pseudo open loop data is computed from available closed loop measurements. In Algorithm~\ref{alg:3step}, $removeTT$ denotes a function which removes the tip-tilt (linear) component of wavefronts, and $\alpha_{TT},B_{TT}$ denote the regularization parameter and sparse matrix \eqref{sparseM}, respectively, for the atmospheric tomography problem corresponding to the TTS only setting. Note that for the computation of the mirror commands, only the measurements of the previous time step are available. Therefore, the mirror updates will have a 2 time step delay. We use an integrator control, where a predefined gain combines the mirror updates of the last two time steps to increase stability, see line 6 of Algorithm~\ref{alg:3step}.

\begin{algorithm}[ht!]
    \caption{Three-step AO with SAFR}\label{alg:3step}
    \begin{algorithmic}[1]
        \setstretch{1.4}
        \Require $i$ (current time step)
        \Statex $s^{(i-1)}$ (measurement vector)
        \Statex $\quad a^{(i)}$ (previous mirror commands) 
        \Statex $\quad \alpha$ (regularization parameter)
        \Statex $\quad \alpha_{TT}$ (regularization parameter for TT-reconstruction) 
        \Statex $\quad \beta$ (smoothing parameter)
        \Statex $\quad gain$ (scalar weight) 
        \Statex $\quad B, \,B_{TT}$ (precomputed, sparse matrices in SAFR)
        
        \State $(\Psi_1,\ldots,\Psi_G) = CuReD(s^{(i-1)})$
        \Comment{Wavefront reconstruction}
        
        \State $(\widetilde{\Psi}_1,\ldots,\widetilde{\Psi}_G)=(\Psi_1,\ldots,\Psi_G)+A (a^{(i)})$ 
        \Comment{POLC}
        
        \State $(\overline{\Psi}_1,\ldots,\overline{\Psi}_{G_{LGS}})=removeTT(\widetilde{\Psi}_1,\ldots,\widetilde{\Psi}_{G_{LGS}})$\Statex\Comment{Tip-Tilt removal for LGS-wavefronts}
        
        \State ${\Phi}^{(i+1)} = SAFR(\overline{\Psi}_1,\ldots,\overline{\Psi}_{G_{LGS}}, \alpha,\beta,B) + SAFR(\widetilde{\Psi}_{G_{LGS+1}},\ldots, \widetilde{\Psi}_G, \alpha_{TT},\beta,B_{TT})$ 
        \Statex\Comment{Atmospheric tomography}
        
        \State $a^{(i+1)} = (1-gain) \cdot a^{(i)} + gain \cdot {\Phi}^{(i+1)} $ \Comment{Integrator control}
        \State\Return $a^{(i+1)}$
    \end{algorithmic}
\end{algorithm}

\subsection{Computational Cost}

One of the main advantages of the proposed SAFR algorithm is the reduced computational cost as well as reduced memory usage compared to standard MVM methods. We examine the cost of the proposed three-step Algorithm~\ref{alg:3step} with SAFR against FEWHA \cite{Yudytskiy_2014,Yudytskiy_Helin_Ramlau_2014} and the COMPASS internal LS-reconstructor (LS).

\begin{itemize}
    \item \textbf{Three-step approach with SAFR:} The computational cost for the three-step method consists of the cost of the computation of pseudo open loop data, wavefront reconstruction via the CuReD, the tip-tilt component removal of LGS wavefronts, the integrator control and atmospheric tomography via SAFR. For $N=M^2$ measurements per guide star, $G$ guide stars and $L$ atmospheric layers, CuReD requires $\mathcal{O}(N)$ Floating Point Operations (FLOPs), while SAFR requires $G$ times the \texttt{fft2}, $L$ times the \texttt{ifft2} and $L\cdot G\cdot N$ computations for the multiplication of the sparse matrix $B$ in \eqref{sparseM}. 
    Pseudo open loop data is computed by applying the atmospheric tomography operator onto the wavefronts. This is a bilinear interpolation, for which we can leverage the possibility of a matrix free implementation with a cost of $\mathcal{O}(N).$
    Note that the domain for the resulting mirror shape updates is predetermined with SAFR to be on an equidistant grid of size $N$ on the domains $[-c_\ell T,c_\ell T]$. In reality, the mirrors might be of different size or spacing, and therefore an additional interpolation step is required. 
    For the soft-real time computations, $N$ SVDs for matrices of size $L\times G$ have to be performed, as well as $NLG$ elementary operations. Computing the SVDs with bidiagonalization and the QR algorithm is $\mathcal{O}(LG\max (L,G))$, and thus the overall computational complexity of the soft-real time computations of SAFR is $\mathcal{O}(N)$.
    The pre-computed coefficients can be stored in the matrix $B$, as defined in \eqref{sparseM}, which amounts to memory storage requirements for a complex matrix of size ${N\times LG}.$ 

    \item \textbf{LS reconstructor:} The LS reconstructor computes a control matrix $R$ via singular value decomposition and pseudo-inversion of the interaction matrix (which maps the DM commands $a$ to the sensor measurements $s$). The reconstruction $R \cdot s^{(i-1)}$ is inserted into the temporal compensator, which is an integrator. The command vector $a$ at time step $i+1$ is computed as 
    \begin{equation}\label{mvm}
        a^{(i+1)} = a^{(i)} + gain \cdot R \cdot s^{(i-1)}\,.
    \end{equation}
    Its computational cost is $\mathcal{O}(N^2)$, required for the matrix-vector multiplication. The inversion of the interaction matrix can be pre-computed in soft-real time. Therefore, it additionally requires memory storage for the control matrix of size ${NL\times NG}.$ 

    \item \textbf{FEWHA:} FEWHA is an iterative method, in which each iteration is based on the discrete wavelet transform, and which has a computational cost of $\mathcal{O}(N).$ It is based on the CG-method, using a relatively small number of iterations $iter$, typically 2-4. \cite{Stadler2022}.  

\end{itemize}
In general, MVM methods have a computational cost of $\mathcal{O}(N^2)$. Using SAFR, the computational demand is mainly caused by  \texttt{fft} leading to a reduced complexity of $\mathcal{O}(N \log N)$. 
In contrast, iterative approaches such as FEWHA or FrIM \cite{Tallon_TallonBosc_Bechet_Momey_Fradin_Thiebaut_2010,Tallon_Bechet_TallonBosc_Louarn_Thiebaut_Clare_Marchetti_2012,Thiebaut_Tallon_2010} yield an asymptotical rate of $\mathcal{O}(N)$.
Note that in practice the matrices $B$ in \eqref{sparseM} and $R$ in \eqref{mvm} are pre-computed, thus, their calculation is not included into the hard real-time costs. Due to, e.g., changes of guide star positions or atmospheric conditions, these matrices may have to be recalculated regularly, which is not necessary for iterative methods such as FEWHA or FrIM.

\begin{table}[]
    \centering
    \resizebox{\columnwidth}{!}{
    \begin{tabular}{|l|c|c|c|}
    \hline
    \multirow{2}{*}{\textbf{Algorithm}}              &      \multirow{2}{*}{\textbf{Operation}}                         & \textbf{Computational cost}& \textbf{memory usage}\\ 
    & & [FLOPs] & [floats]\\
    \hline \hline
    \multirow{7}{*}{SAFR} & \texttt{fft2} & $ G \cdot\mathcal{O}(N \log N)$ & - \\\cline{2-4}
                          &  \texttt{ifft2} & $2L\cdot \mathcal{O}(N \log N)$  &  -  \\\cline{2-4}
                          & Computation of elements of $B$   &   \GRAY{pre-computed,} $\mathcal{O}(N)$      &  $ 2LGN$ \\\cline{2-4}
                          & Multiplication with $B$    &  $L\cdot G\cdot N$     &  -     \\\cline{2-4}
                          &  CuReD & $G\cdot \mathcal{O}(N)$  &  -  \\\cline{2-4}
                          &  Tip-tilt removal & $4G_{LGS}\cdot N$  &  -  \\ \cline{2-4}
                          &  POLC & $L\cdot G\cdot \mathcal{O}(N)$  &  -  \\ \cline{2-4}
                          &  Interpolation to mirror & $L\cdot \mathcal{O}(N)$  &  -  \\\cline{2-4}
                          & \multirow{3}{*}{\textbf{Total cost}} 
                          & $\!\begin{aligned}[t] & (G+L) \cdot\mathcal{O}(N \log N)\\
                          & +(2L\cdot G+L+4)\mathcal{O}(N) \\&\qquad=\mathcal{O}(N\log N)
                            \end{aligned}$   & \multirow{3}{*}{ $ 2LGN$}  \\ \hline
    \multirow{2}{*}{FEWHA}  &  One iteration       & $\mathcal{O}(N)$
                            &  -   \\\cline{2-4}
                            &  \textbf{Total cost} &$iter \cdot \mathcal{O}(N)$ &  -  \\ \hline         
   \multirow{3}{*}{LS}
                    & Inversion of interaction matrix& \GRAY{pre-computed} & $LGN^2$  \\ \cline{2-4}
                    &   Matrix-vector multiplication     &   $\mathcal{O}(N^2)$   &  -   \\\cline{2-4}
                    &   \textbf{Total cost }    &   $\mathcal{O}(N^2)$   & $LGN^2$   \\\hline
    \end{tabular}}
    \caption{List of computational and memory cost of all involved computations.}
    \label{tab:cost}
\end{table}

A summary of the computational costs and memory usage associated with all individual components of the considered methods is provided in Table~\ref{tab:cost}. The memory usage values therein describe the number of floating point numbers (floats) which have to be stored. 
The factor $2$ in memory usage in Table~\ref{tab:cost} for SAFR stems from the fact that the stored matrix is complex, thus requiring $2$ floating point numbers to store one matrix entry.
The computational costs of SAFR and FEWHA mainly differ in the additional $\mathcal{O}(\log N)$ term, compared to the number of iterations in FEWHA. In the numerical experiments we conduct below, we have WFSs with $N=74\times 74$ subapertures. In that specific case, $\log 74^2\approx 3.7$, which roughly coincides in computational cost with the 2-4 iterations that FEWHA usually takes.
Furthermore, SAFR reduces the memory requirements of an MVM method by a factor of $N$, providing a significant improvement. 
Note that we did not consider storage necessities for, e.g., method parameters, which are independent of $N$ and can generally be neglected. 
A precise analysis and comparison of computational cost and memory usage of FEWHA and a general MVM algorithm has been performed in~\cite{Stadler2021}.

In terms of real-time computational demands, it is crucial that algorithms are parallelizable and pipelineable. In this regard, SAFR and MVM methods are advantageous, due to their inherent potential for straightforward parallelization and pipelining. For both methods, matrix-vector multiplications are required, which is a highly parallelizable and pipelineable operation. SAFR, in addition, requires the \texttt{(i)fft2}, for which parallel and pipelined implementations exist, see e.g.~\cite{El-Khashab2002,Groginsky1970,Johnston1983}. In contrast, parallelization and pipelining is only a limited possibility for FEWHA, due to its iterative nature \cite{Stadler2020}.

% % % % % % % % % % % % % % % % 
% % Section - Numerics  % % % %
% % % % % % % % % % % % % % % % 

\section{Numerical Experiments}\label{sec:numerics}

In this section, we evaluate the performance of SAFR against FEWHA and the COMPASS internal LS-reconstructor regarding
reconstruction quality in numerical experiments. The test setting in this paper is motivated by the instrument MORFEO \cite{MORFEO}, which is an AO module of the ELT, operating in MCAO.
For our numerical simulations, we use the software package COMPASS \cite{Ferreira_2018}, which in particular allows simulation of all critical sub-components of an AO system in the context of the ELT. 

% Subsection - Test Configuration and Simulation Environment
\subsection{Test Configuration and Simulation Environment}

The AO system configuration is shown in Table~\ref{tab:general_setting}. We simulate a telescope which gathers light through a primary mirror with a $37$~m diameter and an $11\%$ central obstruction. The ELT optical design consists of three mirrors denoted by M1, M2, and M3 on-axis with two DMs (M4, M5) for performing the AO correction. For MORFEO, two additional DMs (DM1, DM2) inside the instrument are used for wavefront compensation. Note that we assume the Fried geometry with equidistant actuator spacing for all DMs \cite{Fried_77}. Details on the configuration are listed in Table~\ref{tab:DM}. We use a 35-layer atmosphere \cite{Kolb15} for simulation and reconstruct $3$ layers located at the altitudes of the DMs (see Table~\ref{tab:layers}) to avoid the mirror fitting step. 

\begin{table}[ht!]
    \footnotesize
    \renewcommand{\arraystretch}{1.3}
    \begin{minipage}{.49\textwidth}
        \centering  
    	\begin{tabular}{|r|c|}
    		\hline
    		\textbf{Parameter} & \textbf{Value}\\\hline\hline
    		Telescope diameter &  $37$~m\\\hline
    		Central obstruction & $11\%$\\\hline
    		Na-layer height & $90$~km\\\hline
    		Na-layer FWHM & $11.4$~km\\\hline
    		Field of View & $1$~arcmin\\\hline
    		Evaluation criterion & Strehl ratio\\\hline
    		Evaluation wavelength & K band ($2200$~nm) \\\hline
    	\end{tabular}
     	\caption{General system parameters for the simulation.}
      	\label{tab:general_setting}
    \end{minipage}
    \begin{minipage}{.49\textwidth}
        \centering
    	\begin{tabular}{|r|c|c|c|}
    		\hline
    		& \textbf{M4} & \textbf{DM1} & \textbf{DM2}\\\hline\hline
    		Actuators & $75\times 75$ &  $47\times 47$ & $37\times 37$\\\hline
    		Altitude & $0$~km & $4$~km & $12.7$~km\\\hline
    		Spacing & $0.5$~m & $1$~m & $1$~m\\\hline
    	\end{tabular}
     	\caption{DM configuration.}
       	\label{tab:DM}
    \end{minipage}
\end{table}

\begin{figure}[ht]
\centering
\includegraphics[width=.7\textwidth]{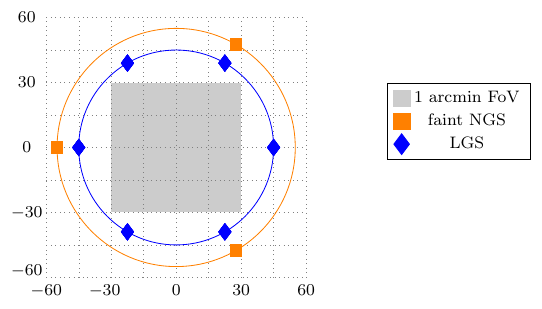}
	\caption{Different configurations of faint NGSs (orange) and LGSs (blue). The $1$ arcmin FoV is marked in gray.}
	\label{fig:MCAO_asterism}
\end{figure}

In the standard MORFEO setting (Fig.~\ref{fig:MCAO_asterism}), $6$ LGS are positioned in a circle of $90$~arcsec diameter. We account for the tip-tilt effect via removing the planar component of LGS wavefronts using $3$ faint NGS (TTS) located in a circle of $110$~arcsec diameter \cite{Gilles_Ellerbroeck_2008}. We model the sodium layer at which the LGS beam is scattered via a Gaussian random variable with mean altitude $H = 90$~km and FWHM (full width at half maximum) of the sodium density profile of $11.4$~km. To each LGS, a SH WFS with $74\times74$ subapertures is assigned. The faint NGS for tip-tilt removal are equipped with $2\times2$ SH WFS. The noise induced by the detector read-out is simulated as $3.0$ electrons per pixel and frame. For more details see Table~\ref{tab:WFS}.

\begin{table}
\footnotesize
\renewcommand{\arraystretch}{1.3}
\begin{minipage}{.49\textwidth}
    \centering
	\begin{tabular}{|c|c|c|}
		\hline
		\textbf{Layer} & \textbf{Altitude} & \textbf{Strength}\\\hline\hline
		$1$ & $0$~m & $0.75$\\\hline
		$2$ & $4000$~m & $0.15$\\\hline
		$3$ & $12700$~m & $0.1$\\\hline
	\end{tabular}
 	\caption{Layer configuration.}
  	\label{tab:layers}
\end{minipage}
\begin{minipage}{.49\textwidth}
    \centering  
	\begin{tabular}{|r|c|c|c|}
		\hline
		& \textbf{LGS-WFS} & \textbf{TT-WFS}\\\hline\hline
		Type & SH WFS & SH WFS\\\hline
		Subap. & $74\times 74$ & $2\times 2$\\\hline
		Wavelength & $589$~nm & $1650$~nm\\\hline
	\end{tabular}
 	\caption{WFS configuration.}
  	\label{tab:WFS}
\end{minipage}
\end{table}

% \begin{figure}[ht]
% \centering
%  \begin{tikzpicture}[scale=1.1,font=\footnotesize]
% 	\fill[gray!40!white] (2,2) rectangle (4,4);
% 	\draw [very thin, dotted, gray, step=0.5] (0.9999,0.9999) grid (5,5);
% 	\draw (0.7,1.3) node[below] {$-60$};
% 	\draw (0.7,2.2) node[below] {$-30$};
% 	\draw (0.7,3.2) node[below] {$0$};
% 	\draw (0.7,4.2) node[below] {$30$};
% 	\draw (0.7,5.2) node[below] {$60$};
% 	\draw (1,1) node[below] {$-60$};
% 	\draw (1.9,1) node[below] {$-30$};
% 	\draw (3,1) node[below] {$0$};
% 	\draw (4,1) node[below] {$30$};
% 	\draw (5,1) node[below] {$60$};
% 	\draw[blue] (3,3) circle (1.5);
% 	\draw[orange] (3,3) circle (1.8333);
% 	\foreach \x in {2,2.5,3,3.5,4}
%     \draw [blue] plot [only marks, mark size=3.5, mark=diamond*] coordinates {(4.5,3) (3.75,4.299) (2.25,4.299) (1.5,3) (2.25,-1.299+3) (3.75,-1.299+3)};
%     \draw [orange] plot [only marks, mark size=2.5, mark=square*] coordinates {(3.9167,4.5877) (-1.8333+3,3) (3.9167,-1.5877+3)};
%     \hspace{0.5cm}
% 	\begin{customlegend}[
% 	legend entries={
% 		$1$ arcmin FoV,
%         faint NGS,
% 		LGS,
% 	},
% 	legend style={at={(8,4)}}]
% 	\addlegendimage{only marks, mark=square*, color=gray!40!white, mark size=4}
%     \addlegendimage{only marks, mark=square*, color=orange, mark size=4}
% 	\addlegendimage{only marks, mark=diamond*, color=blue, mark size=5}
% 	\end{customlegend}
% 	\end{tikzpicture}
% 	\caption{Different configurations of faint NGSs (orange) and LGSs (blue). The $1$ arcmin FoV is marked in gray.}
% 	\label{fig:MCAO_asterism}
% \end{figure}

To validate the reconstruction quality, we use long exposure (LE) and short exposure (SE) Strehl ratio into certain directions \cite{Roddier1999}. The Strehl ratio is defined as the ratio between the maximum of the real energy distribution of incoming light in the image plane $I(x,y)$ over the hypothetical distribution $I_D(x,y)$, which stems from the assumption of diffraction-limited imaging, i.e.,
\begin{equation*}
    \text{SR}:= \frac{\max_{(x,y)}I(x,y)}{\max_{(x,y)}I_D(x,y)} \,.
\end{equation*}
By definition, the Strehl ratio is between $0$ and $1$ and frequently given in percent. A Strehl ratio of $1$ means that the influence of the atmosphere has been removed from the observation. For its numerical evaluation, the Marechal criterion is used \cite{Roddier1999}. We simulate $10000$ time steps, corresponding to a $20$ seconds interval.

\begin{table}
    \centering
    \begin{tabular}{|l|c|c|c|}
    \hline
                          &     Parameter                          & \textbf{lowflux} & \textbf{highflux} \\ \hline \hline
    \multirow{4}{*}{\rotatebox[origin=c]{90}{SAFR}} & $\alpha$ & 0.005 &0.005 \\\cline{2-4}
                          &  $\alpha_{TT}$ &  0.001  &      0.01    \\\cline{2-4}
                          & $\beta  $          &   1.5      &    1.5\\\cline{2-4}
                          &  $gain$        &   0.5      &    1  \\\hline
    \multirow{5}{*}{\rotatebox[origin=c]{90}{FEWHA}}     &   $iterations$     &  2-4     &  2-4   \\\cline{2-4}
                          &           $\alpha$                    &    100     &  250        \\\cline{2-4}
                          &          $\alpha_\eta$                     &   0.2      &     0.2     \\\cline{2-4}
                          &       $\alpha_J$                      &    $10^6$   &    $10^6$      \\\cline{2-4}
                          &        $gain$                       &   0.6      &   0.8      \\\hline
                          \rotatebox[origin=c]{90}{LS} & $gain$ & 0.5 & 0.8  \\ \hline
    \end{tabular}
    \caption{Method parameters used in the numerical experiments.}
    \label{tab:parameters}
\end{table}

All method specific parameters for SAFR, FEWHA, and the LS algorithm were optimized in COMPASS a-priori. The parameters used in the numerical simulations shown in this paper are summarized in Table~\ref{tab:parameters}, a precise description of the origin and effect of the FEWHA parameters can be found in~\cite{Stadler2022}. For the SAFR Algorithm~\ref{alg:SAFR_hrt},~\ref{alg:SAFR_srt}, we used $\beta=1.5$ for the smoothing parameter, which is slightly smaller than the choice $\beta=11/6$ corresponding to the expected smoothness of a typical atmosphere \cite{Kolmogorov}. This was done to avoid the commonly observed oversmoothing effect inherent in regularization methods with Sobolev norm penalty terms \cite{Hubmer_Sherina_Ramlau_2023,Ramlau_Teschke_2004_1}.

% Subsection - Numerical Results
\subsection{Numerical Results}

% \begin{figure}[hp]
%     \centering
    
%     % First row
%     \begin{subfigure}[b]{\textwidth}
%         \centering
%         \includegraphics[width=0.48\textwidth]{Figures/lowflux_mixed_r0_0097_avSE.png}
%         \includegraphics[width=0.48\textwidth]{Figures/lowflux_mixed_r0_0097_LEsep.png}
%         \caption{Bad seeing conditions, $r_0=0.097\,m$.}
%         \label{fig:lowflux_badseeing}
%     \end{subfigure}
    
%     \vspace{1em} % Adjust spacing between rows
    
%     % Second row
%     \begin{subfigure}[b]{\textwidth}
%         \centering
%         \includegraphics[width=0.48\textwidth]{Figures/lowflux_mixed_r0_0157_avSE.png}
%         \includegraphics[width=0.48\textwidth]{Figures/lowflux_mixed_r0_0157_LEsep.png}
%         \caption{Medium seeing conditions, $r_0=0.157\,m$.}
%         \label{fig:lowflux_middleseeing}
%     \end{subfigure}
    
%     \vspace{1em} % Adjust spacing between rows
    
%     % Third row
%     \begin{subfigure}[b]{\textwidth}
%         \centering
%         \includegraphics[width=0.48\textwidth]{Figures/lowflux_mixed_r0_0234_avSE.png}
%         \includegraphics[width=0.48\textwidth]{Figures/lowflux_mixed_r0_0234_LEsep.png}
%         \caption{Good seeing conditions, $r_0=0.234\,m$.}
%         \label{fig:lowflux_goodseeing}
%     \end{subfigure}
    
%     \caption{\RED{SE Strehl ratio averaged over the field of view vs.\ time} (left) and LE Strehl ratio vs. separation (right) after 10,000 time steps for different seeing conditions (different Fried parameter $r_0$) in the lowflux setting with 500 photons per subaperture per frame.}
%     \label{fig:lowflux_results}
% \end{figure}

\begin{figure}[hp]
    \centering
    \includegraphics[width=\textwidth]{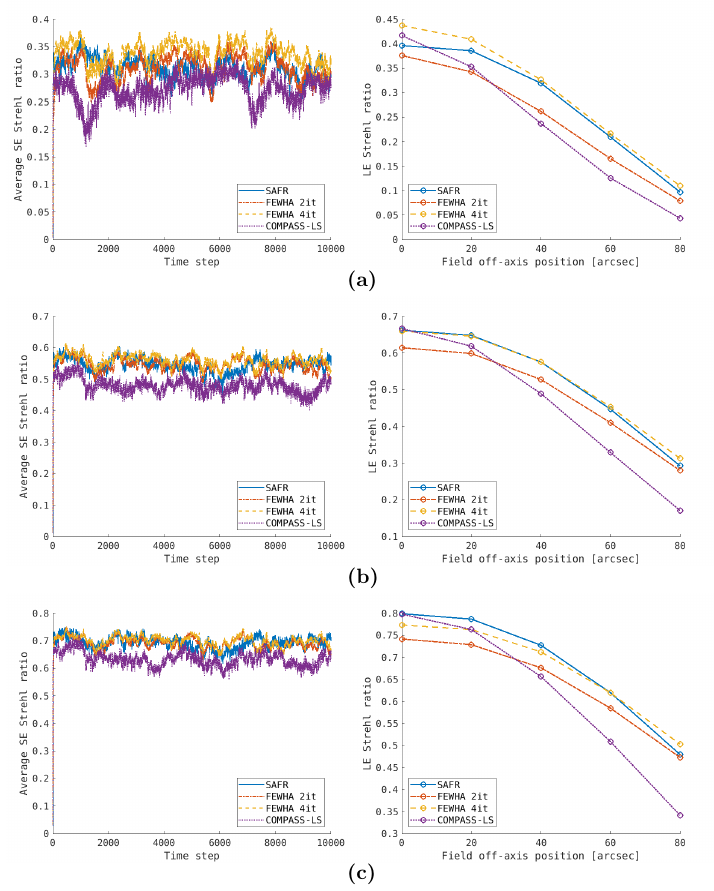}
    \caption{SE Strehl ratio averaged over the field of view vs.\ time (left) and LE Strehl ratio vs. separation (right) after 10,000 time steps for different seeing conditions in the lowflux setting with 500 photons per subaperture per frame: (a) Bad seeing conditions, $r_0=0.097\,\mathrm{m}$. (b) Medium seeing conditions, $r_0=0.157\,\mathrm{m}$. (c) Good seeing conditions, $r_0=0.234\,\mathrm{m}$.}
    \label{fig:lowflux_results}
\end{figure}

\begin{figure}[ht]
    \centering
    % [trim={left bottom right top},clip]
    \includegraphics[width=\textwidth,trim={0 0 0 0},clip]{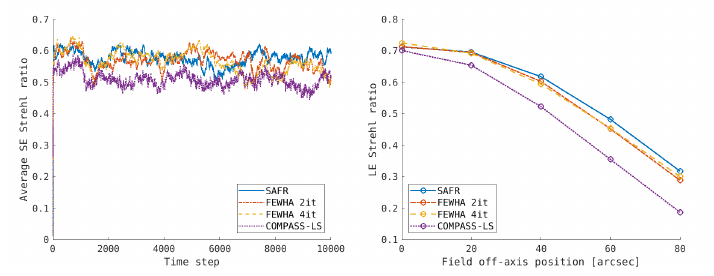}
    \caption{SE Strehl ratio averaged over the field of view vs.\ time (left) and LE Strehl ratio vs. separation (right) after 10,000 time steps for medium seeing conditions $r_0=0.157m$ in the highflux setting with 10,000 photons per subaperture per frame.}
    \label{fig:highflux_results}
\end{figure}

First, we consider a lowflux setting with 500 photons per subaperture per frame for all guide stars, and analyze the performance for different seeing conditions, i.e., for several values of the Fried parameter $r_0$. The Fried parameter indicates the medium length of turbulence in the atmosphere, with larger values of $r_0$ corresponding to better seeing conditions, for which generally better results are expected than for lower values of $r_0$. The results are summarized in Fig.~\ref{fig:lowflux_results}. The left plots in Fig.~\ref{fig:lowflux_results} show the average (over the FoV) SE Strehl ratio over time, while the right plots show the LE Strehl ratio over separation, i.e., distance from the center. We observe that SAFR (blue, solid line) qualitatively outperforms the LS reconstructor (purple, dotted line) in almost every case, with an exception in the very center direction in the $r_0=0.097m$ case, i.e., bad seeing conditions in Fig.~\ref{fig:lowflux_results}a. Particularly in outer directions, SAFR yields a much higher LE Strehl ratio than the LS reconstructor. Furthermore, we compared FEWHA with $2$ (red, dash-dotted line) and $4$ (yellow, dashed line) CG iterations. In~\cite{RaSt2021, Stadler2021,Stadler2020}, it has been shown that the reconstruction quality of FEWHA does not increase significantly for more than 4 iterations. Furthermore, it is worth to note that - with current computational standards - FEWHA with $2$ iterations fulfills the real-time requirements of the ELT, while FEWHA with $4$ iterations does not. Again, SAFR outperforms FEWHA with $2$ iterations in every case and direction, only FEWHA with $4$ iterations competes qualitatively with SAFR. For medium seeing conditions ($r_0=0.157m$, Fig.~\ref{fig:lowflux_results}b), the LE Strehl is very similar for FEWHA and SAFR. For good seeing conditions ($r_0=0.234m$, Fig.~\ref{fig:lowflux_results}c), SAFR generally performs better, while FEWHA with $4$ iterations performs best for bad seeing conditions ($r_0=0.097m$, Fig.~\ref{fig:lowflux_results}a). We expect that the difference in performance behavior between SAFR and FEWHA for different seeing conditions can be reasoned with the choice of the regularization parameter. While FEWHA has been developed and optimized for several years already, in-depth parameter studies for our SAFR approach have not been performed so far. However, similar levels of robustness with respect to noise are expected for both methods, since they employ two closely related Tikhonov-type regularization approaches.

For highflux data, i.e., $10,000$ photons per subaperture per frame, we only considered the case of medium seeing conditions $r_0=0.157\,m$ in Fig.~\ref{fig:highflux_results}. Here, the results of the considered methods in terms of LE Strehl ratio are generally closer together, where, again, FEWHA and SAFR outperform the LS reconstructor. Notably, more distant from the center, SAFR performs even better than FEWHA in this case. Finally, note that it is a well-known issue for MCAO systems that the quality of the AO correction degrades further away from the center, see e.g.~\cite{Ramlau_Stadler_2025}.

% % % % % % % % % % % % %
% Section - Conclusion  %
% % % % % % % % % % % % %
\section{Conclusion}\label{sec:conclusion}

In this paper, we proposed the SAFR algorithm for the solution of the atmospheric tomography problem, which is part of three-step adaptive optics approaches. We analyzed its computational cost and investigated its performance in an MCAO setting resembling the ELTs MORFEO instrument in numerical simulations. The results obtained within the simulation environment COMPASS showed that in comparison with both the COMPASS internal LS reconstructor and FEWHA, an iterative, matrix-free reconstructor, SAFR qualitatively competes with FEWHA and outperforms the LS reconstructor (an MVM algorithm). In contrast to FEWHA, parallelization and pipelining can be implemented more easily for SAFR, which only relies on standard operations such as matrix multiplication and the discrete Fourier transform. SAFR directly provides a regularized solution to the atmospheric tomography problem without the need for iterations. Furthermore, we found a major improvement with SAFR in terms of both computational expense and memory usage compared to the LS reconstructor in COMPASS.
Finally, note that although the analysis of its hard real-time computational complexity showed that SAFR and FEWHA are similar, this does not necessarily imply that this also holds for their actual run-time. While there are efficient implementations of FEWHA on real-time hardware, showing that it meets the requirements for the MORFEO instrument using 2 iterations \cite{Stadler2022}, a parallelized and pipelined version of SAFR is not yet available and subject of future work.

% % % % % % % % % % % %
% Section - Appendix  %
% % % % % % % % % % % %
\appendix

% Appendix A - Derivation of the SAFR Algorithm
\section{Derivation of the SAFR Algorithm}\label{appendix}

In this appendix, we provide the mathematical details of the derivation of the SAFR algorithm. First, using the discretization grids 
    \begin{equation*}
        (x_p^\ell,y_q^\ell)=(-c_{\ell}T+2c_{\ell}T \cdot p/M,-c_{\ell}T+2c_{\ell}T \cdot q/M),\qquad  \text{for} \quad p,q=0,\ldots,M-1 \,,
    \end{equation*}
and
    \begin{equation*}
        (x_p,y_q)=(-T+2T \cdot p/M,-T+2T \cdot q/M),\qquad  \text{for} \quad p,q=0,\ldots,M-1 \,,
    \end{equation*}
we obtain 
    \begin{equation}\label{wjk_dft_app}
        w_{jk}(x_p,y_q)=(-1)^{j+k}\frac{1}{2T}e^{2\pi i(jp+kq)/M} \,,
    \end{equation}
and
    \begin{equation}\label{wjkl_dft_app}
        w_{jk,\ell}^{(0)}(x_p^\ell,y_q^\ell)=(-1)^{j+k}\frac{\gamma_\ell^{1/2}}{2c_\ell T}e^{2\pi i(jp+kq)/M} \,.
    \end{equation}
Next, we truncate the infinite sums in \eqref{SVD_reg_algo} symmetrically at $\pm m$ with $M=2m+1$ (implicitly assuming that $M$ is odd) to obtain the approximation
    \begin{equation*}
        \left(\phi_\alpha^\delta\right)_{\ell}(x_p^\ell,y_q^\ell)\approx
        \sum_{j,k=-m}^{m} b_{jk,\ell} w_{jk,\ell}^{(0)}(x_p^\ell,y_q^\ell) \,.
    \end{equation*}
To align this finite sum with the indices used in the definition of the DFT, we use
    \begin{equation}\label{helper_01_app}
      \left(\phi_\alpha^\delta\right)_{\ell}(x_p^\ell,y_q^\ell)
      \approx
      \sum_{j,k=0}^{M-1} b_{n(j)n(k),\ell} w_{n(j)n(k),\ell}^{(0)}(x_p^\ell,y_q^\ell)\;.
    \end{equation}
Furthermore, due to 
    \begin{equation*}
        w_{jk,\ell}^{(0)}(x_p^\ell,y_q^\ell)=w_{(j+r\cdot M) (k +r\cdot M),\ell}^{(0)}(x_p^\ell,y_q^\ell)\;,\qquad \forall\, r \in \Z\;,
    \end{equation*}
there holds $w_{n(j)n(k),\ell}(x_p^\ell,y_q^\ell)=w_{jk,\ell}(x_p^\ell,y_q^\ell)$. Hence, together with \eqref{helper_01_app}, we obtain
    \begin{equation*}%\label{idft_approximation_app}
        \left(\phi_\alpha^\delta\right)_{\ell}(x_p^\ell,y_q^\ell)\approx
        \sum_{j,k=0}^{M-1} b_{n(j)n(k),\ell} w_{jk,\ell}^{(0)}(x_p^\ell,y_q^\ell) \\
        \overset{\eqref{wjkl_dft_app}}{=} \frac{\gamma_\ell^{1/2}M^2}{2c_\ell T}\texttt{IDFT2}(b_{\ell})_{p+1,q+1} \;,
    \end{equation*}
with the vector $b_{\ell} := ((-1)^{j+k}b_{n(j)n(k),\ell})_{j,k=0}^{M-1}$. Similarly, we are able to approximate the coefficient vectors $\varphi^\delta_{jk}$ by
    \begin{equation}\label{coeff_approximation_app}
    \begin{aligned}
        (\varphi^\delta_{n(j)n(k)})_g
        &\overset{\hphantom{\eqref{wjk_dft_app}}}{=}\skp{\varphi_g^\delta,w_{n(j)n(k)}}_{L_2(\Omega_T)}\approx\frac{(2T)^2}{M^2}\sum_{p,q=0}^{M-1} \varphi_g^\delta(x_p,y_q) \overline{w}_{n(j)n(k)}(x_p,y_q)\\&\overset{\hphantom{\eqref{wjk_dft_app}}}{=} \frac{(2T)^2}{M^2}\sum_{p,q=0}^{M-1}\varphi_g^\delta(x_p,y_q) \overline{w}_{j,k}(x_p,y_q) \\&\overset{\eqref{wjk_dft_app}}{=} (-1)^{j+k}\frac{2T}{M^2}\texttt{DFT2} (\Psi^\delta_g)_{j+1,k+1} \,, \qquad \text{for} \quad g=1,\ldots G \,.
    \end{aligned}     
    \end{equation}
Note that \eqref{coeff_approximation_app} amounts to using a right rectangular quadrature rule for the integral in the $L_2(\Omega_T)$ inner product. 
Now, if we define the coefficients
    \begin{equation*}
        b_{jk,\ell g}:=\frac{\gamma_\ell^{1/2}}{c_\ell}\left(1+\tau_{\ell,T}|(jk)|^{2}\right)^{-\beta/2}\left(\mathcal{R}_{\alpha}(A_{jk})\right)_{\ell g}\,,
    \end{equation*}
there holds
    \begin{equation*}%\label{bjkl_approximation_app}
        b_{n(j)n(k),\ell} \overset{\eqref{def_bjkl},\eqref{coeff_approximation_app}}{\approx} (-1)^{j+k}\frac{2c_\ell T}{\gamma_\ell^{1/2}M^2}\sum_{g=1}^G b_{n(j)n(k),\ell g} \cdot \texttt{DFT2}(\Psi_g^\delta)_{j+1,k+1}\;.
    \end{equation*}
Hence, combining the above expressions, we obtain
    \begin{equation*}
        \left(\phi_\alpha^\delta\right)_{\ell}(x_p^\ell,y_q^\ell)\approx
        \frac{\gamma_\ell^{1/2}M^2}{2c_\ell T}\texttt{IDFT2}(b_{\ell})_{p+1,q+1} \,,
    \end{equation*}
with 
    \begin{equation*}
        b_{\ell} = \left(\frac{2c_\ell T}{\gamma_\ell^{1/2}M^2}\sum_{g=1}^G \frac{\gamma_\ell^{1/2}}{c_\ell}\left(1+\tau_{\ell,T}|(jk)|^{2}\right)^{-\beta/2}\left(\mathcal{R}_{\alpha}(A_{jk})\right)_{\ell g} \cdot \texttt{DFT2}(\Psi_g^\delta)_{j+1,k+1}\right)_{j,k=0}^{M-1} \,,
    \end{equation*}
which now yields \eqref{idft_approximation} and \eqref{bjkl_approximation}.

% % % % % % % % % % %
% Section - Support %
% % % % % % % % % % %
\section*{Acknowledgment}

The authors thank Dr.\ Daniel Jodlbauer for assistance with COMPASS. RR and SH were funded in part by the Austrian Science Fund (FWF) SFB 10.55776/F68 ``Tomography Across the Scales'', project F6805-N36 (Tomography in Astronomy). For open access purposes, the authors have applied a CC BY public copyright license to any author-accepted manuscript version arising from this submission. Furthermore, the authors were supported by the NVIDIA Corporation Academic Hardware Grant Program. LW and BS are partially supported by the State of Upper Austria. 

\sloppy
\printbibliography    
\fussy

%===============================================================================================

\end{document}

%% file: header.tex
\usepackage{hyperref}
\usepackage{url}
\usepackage{bbm}
\usepackage[backend=bibtex, style=numeric,
    giveninits=true, maxbibnames=5]{biblatex}
\AtBeginBibliography{\footnotesize}
\usepackage{amsthm}
\usepackage{amsmath}
\usepackage{amssymb}
\usepackage{graphics}
\usepackage{enumitem}
\usepackage{color}

\usepackage{mathtools}
\usepackage[usenames,dvipsnames]{xcolor}
\usepackage{gensymb}
\usepackage{tabularx}
\usepackage{multirow}
\usepackage{aligned-overset}
\usepackage{harpoon}
\usepackage{overpic}
\usepackage{pbox}
\usepackage{algorithm}
\usepackage{algpseudocode}

\usepackage{setspace}
\usepackage{subcaption}
\usepackage[left=3cm, right=3cm, top=3cm, bottom=4cm]{geometry}
\usepackage{tikz}
\usetikzlibrary{patterns}
\numberwithin{equation}{section}
\allowdisplaybreaks
\definecolor{myblue}{RGB}{0, 0, 0}
\definecolor{mygray}{gray}{0.6}
\newcommand{\GRAY}[1]{\textcolor{mygray}{#1}}

\renewbibmacro*{doi+eprint+url}{%
    \iffieldundef{doi}{%
        \iffieldundef{eprint}{%
            \iffieldundef{url}{%
            }{%
                \usebibmacro{url+urldate}%
            }%
        }{%
            \usebibmacro{eprint}%
        }%
    }{%
        \printfield{doi}%
    }%
}

\hypersetup{
    unicode=false,          % non-Latin characters in Acrobat’s bookmarks
    pdftoolbar=true,        % show Acrobat’s toolbar?
    pdfmenubar=true,        % show Acrobat’s menu?
    pdffitwindow=false,     % window fit to page when opened
    pdfstartview={FitH},    % fits the width of the page to the window
    pdftitle={},    % title
    pdfauthor={Lukas Weissinger},     % author
    pdfsubject={paper},   % subject of the document
    pdfcreator={},   % creator of the document
    pdfkeywords={}, % list of keywords
    pdfnewwindow=true,      % links in new PDF window
    colorlinks=true,       % false: boxed links; true: colored links
    linkcolor=myblue,          % color of internal links (change box color with linkbordercolor)
    citecolor=myblue,        % color of links to bibliography
    linkbordercolor={0 0 0},
    filecolor=black,      % color of file links
    urlcolor=black           % color of external links
}

\if@twoside
\else 
\fi

\numberwithin{equation}{section}
\counterwithin{figure}{section}
\counterwithin{table}{section}

\theoremstyle{plain}% default

\theoremstyle{definition}

\theoremstyle{remark}

\newcommand{\skp}[1]{\left\langle\,#1\,\right\rangle}
\newcommand{\spr}[1]{\left\langle\,#1\,\right\rangle}
\newcommand{\kl}[1]{\left(#1\right)}

\definecolor{aog}{rgb}{0.0, 0.5, 0.0}

\newcommand{\R}{\mathbb{R}} 
\newcommand{\C}{\mathbb{C}}

\newcommand{\Z}{\mathbb{Z}}

\newcommand{\LGS}{{\operatorname{LGS}}}

\usepackage{pgfplots}
\usepackage{pgfplotstable,booktabs}

\def\addlegendimage{\csname pgfplots@addlegendimage\endcsname}
\pgfkeys{/pgfplots/number in legend/.style={
		/pgfplots/legend image code/.code={
			\node at (0.295,-0.0225){#1};
		},
	},
}